\documentclass[fleqn,12pt,twoside]{article}
\usepackage{espcrc1}
\usepackage{amssymb}

\usepackage{graphicx}
\usepackage[figuresright]{rotating}

\begin{document}

\title{Strangeness Production in $pp$ and $pn$ Reactions at COSY%
}

\author{M.~B\"uscher
\address{Institut f\"ur Kernphysik, 
         Forschungszentrum J\"ulich, 52425 J\"ulich} }
\maketitle

\begin{abstract}
The COoler SYnchrotron COSY-J\"ulich delivers phase-space cooled,
polarized proton and deuteron beams with momenta up to
$p{=}3.65$~GeV/c.  Various experiments on hadron-induced strange\-ness
production on proton, neutron and nuclear targets have been carried
out.  Here we report about recent results on associated strangeness
production in $pp{\to}KYN$ ($Y{=}\Lambda,\, \Sigma$) reactions, on
$K^+$-production in $pn$ collisions, and on $K\bar K$-pair production
in $pp$ interactions.  We also briefly discuss possible measurements
to disentangle the parity of the recently discovered pentaquark state
$\Theta^+$, the spin dependence of the $YN$ interaction, as well as
planned experiments which aim at the determination of the $a_0$-$f_0$
mixing matrix element, a quantity which is believed to be sensitive to
the nature of the light scalar mesons $a_0$/$f_0$(980).
\end{abstract}

\section{Experimental facilities at COSY}
COSY-J\"ulich~\cite{cosy} provides electron or stochastically cooled,
polarized proton and deuteron beams with momenta up to $p{=}
3.65$~GeV/c, corresponding to beam energies of $T_p{=}2.83$ and
$T_d{=}2.23$ GeV, respectively. Various experimental facilities at
internal or external target positions can be used for the study of
$K$-meson production in hadronic interactions.

In measurements with thin and windowless internal targets, secondary
processes of the produced mesons can be neglected and, simultaneously,
sufficiently high luminosities are obtained. For the $pp$ and $pn$
measurements at the ANKE and COSY-11 facilities a cluster-jet
target~\cite{clustertarget} providing areal densities of up to $\sim
5\times 10^{14}$~cm$^{-2}$ has been used (with H$_2$ or D$_2$ as
target material). With proton beam intensities of a few $10^{10}$
luminosities of $\mathcal{L} \gtrsim 10^{31}\, \mathrm{cm^{-2}s^{-1}}$
have been achieved.

The ANKE spectrometer~\cite{ANKE_NIM} comprises three dipole magnets
D1--D3 which separate forward-emitted charged reaction products from
the circulating COSY beam and allow one to determine their emission
angles and momenta.  With a gap height of ${\sim} 20$~cm, the
spectrometer dipole D2 provides a large angular acceptance of up to
$\pm 7^{\circ}$ vertically and $\pm 12^{\circ}$ horizontally, which is
particularly advantageous for correlation measurements with threshold
kinematics.  The layout of ANKE, including detectors and the DAQ
system, has been optimized and used to study $K^+$-spectra from $pA$
collisions at beam energies down to $T_p{=} 1.0$~GeV
($p{=}1.70$~GeV/c)~\cite{1.0GeV_PRL,anke_K-potentials,anke_inclusive_K},
thus far below the free nucleon-nucleon threshold at
$T_{NN}{=}1.58$~GeV. This is a very challenging task because of the
small $K^+$-production cross sections, \textit{e.g.}\ 39~nb for $p\,$C
collisions at 1.0~GeV.  In subsequent experiments ANKE has been used
to measure $K^+$-mesons in coincidence with protons and deuterons from
$pn{\to} p K^+ X$ (Sect.~\ref{sec:pn}), $pp{\to}d K^+\bar K^0$
(Sect.~\ref{sec:KK_data}), and $p^{12}\mathrm{C}{\to}p/d\, K^+
X$~\cite{K_corr_EPJA} reactions.

The COSY-11 experiment~\cite{cosy-11} has been designed to study meson
production processes very close to the corresponding production
thresholds. Positively charged ejectiles leaving the interaction
region at forward angles are separated from the circulating COSY beam
and are momentum analyzed in one of the C-shaped ring dipole magnets
downstream of the target region. The 4-momenta of protons and
$K^+$-mesons are measured, leaving the non-identified hyperon
($pp{\to} pK^+Y$
reactions~\cite{cosy-11-pklambda1,cosy-11-pklambda2,cosy-11-lambdanfsi,cosy-11-lambdasigma})
or the $K^-$-meson ($pp{\to} ppK^+K^-$~\cite{cosy-11-KK}) to be
identified by a missing mass analysis.  In case of the $ppK^+K^-$
final state (see Sect.~\ref{sec:KK_data}) an additional silicon pad
detector, mounted inside of the dipole magnet, has been utilized to
measure the hit position of the $K^-$ candidates and to further reduce
the background from other reaction channels.

At the external target positions of COSY a liquid hydrogen (LH$_2$) or
deuterium (LD$_2$) target is being used with a size of typically a few
mm$^3$.  The diameter of the electron cooled and stochastically
extracted beam can be less than 2~mm with a very small beam halo,
while the momentum spread $\Delta p/p$ is less than $10^{-4}$.  The
small interaction region permits to install vertex counters very close
to the interaction point for precise track reconstruction of charged
ejectiles.

The external experiment COSY-TOF~\cite{cosy-tof} is a wide angle, non
magnetic spectrometer with various start and stop detector components
for time-of-flight measurements, combining high efficiency and
acceptance with a moderate energy and momentum resolution. The system
allows one to completely reconstruct the 4-vectors of all final
particles from $pp{\to}KYN$ events including the determination of the
delayed decay vertices of $\Lambda$- and $\Sigma$- hyperons as well of
the $K_S^0$.  Associated strangeness production has been measured in
the reactions $pp{\to}pK^+\Lambda$, $pK^+\Sigma^0$, $pK^0\Sigma^+$,
and $nK^+\Sigma^+$ at beam momenta between 2.5 and 3.3 GeV/c
(Sect.~\ref{sec:associated}). First tests on the corresponding
reactions in $pn$ interactions have been performed with
the LD$_2$ target.

The MOMO experiment~\cite{momo} focuses on near threshold meson-pair
production via the reactions $pd {\to} ^3\mathrm{He}\,
\pi^+\pi^-/K^+K^-$ (Sect.~\ref{sec:KK_data}). The setup comprises a 
scintillating-fiber meson detector near the external LD$_2$ target
with a opening angle $\pm45^{\circ}$, and the high resolution magnetic
spectrometer BIG KARL which is used for $^3$He identification.

For future double polarization experiments a frozen spin NH$_3$ target
(TOF)~\cite{fst} and a polarized internal H$_2$/D$_2$ gas target
utilizing a storage cell (ANKE)~\cite{abs} are presently being
developed.

\section{Associated strangeness production at COSY-11 and COSY-TOF}
\label{sec:associated}
Investigation of associated strangeness production close to threshold
may provide insight into the dynamics of the production processes,
{\em e.g.\/} the role of $N^{\star}$-resonances or the effect of
hyperon-nucleon final-state interaction (FSI).  Total $pp{\to}pKY$
cross sections from COSY-11~\cite{cosy-11-lambdasigma} and
COSY-TOF~\cite{tof_pklambda} at excitation energies $Q<150$~MeV have
been described within various approaches, {\em e.g.\/} the $K$-$\pi$
exchange models of Laget~\cite{laget_tof} and
Sibirtsev~\cite{sibirtsev_tof} as well as resonance-model calculations
of Dillig~\cite{dillig_tof}, Tsushima~\cite{tsushima_tof} and
Shyam~\cite{shyam_tof}. In order to discriminate between these models
precise data, covering full phase space, are needed for different
isospin channels. Spin observables should also be measured, in
particular the polarization of the hyperon, which can be extracted via
its self-analyzing weak decay.

\subsection{\boldmath Production mechanisms in $pp$ collisions}
\label{sec:mechanisms}

\begin{figure}[htb]
  \begin{center}
  \vspace*{-5mm}
  \resizebox{12cm}{!}{\includegraphics[scale=1]{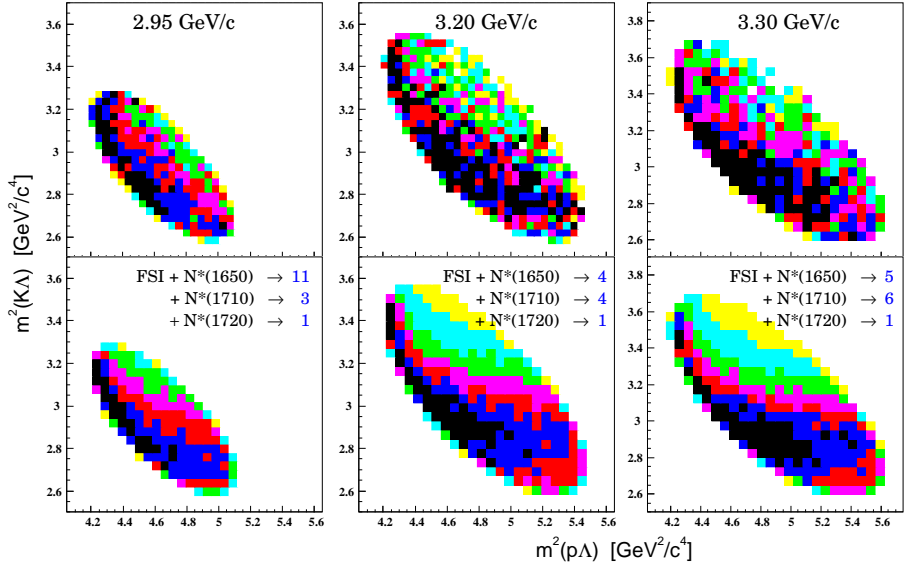}}
  \vspace*{-10mm}
  \caption{Dalitz plots measured at COSY-TOF for the reaction
  $pp{\to}pK^+\Lambda$ (upper) at beam momenta of 2.95 GeV/c,
  3.20 GeV/c and 3.30 GeV/c~\cite{tof_annrep2003}. 
  The lower plots depict the results of calculations with the model of
  Sibirtsev~\cite{sibirtsev_tof} where the strengths of
  the amplitudes of the various resonances have been adjusted to the
  data as indicated.}
  \label{fig:tof_pklambda}
  \end{center}
\end{figure}

Dalitz plot analyses of data from COSY-TOF are a powerful tool to
investigate strange\-ness-production mechanisms. This is demonstrated
in Fig.~\ref{fig:tof_pklambda} where the data~\cite{tof_annrep2003}
are compared with calculations performed with a model of
Sibirtsev~\cite{sibirtsev_tof}, in which contributions from non-resonant
meson exchanges are included coherently with those from the
$N^{\star}$(1650, 1710, 1720)-resonances and the $p\Lambda$-FSI. The
strength of each effect can be adjusted separately to find the best
agreement between data and model.

Whereas at 2.85 GeV/c only the $N^{\star}$(1650) is relevant (not
shown here), a significant contribution from the $N^{\star}$(1710) is
needed at 2.95 GeV/c. At 3.20 GeV/c the amplitudes of both
resonances are equal within the precision of the analysis, and at 3.30
GeV/c the $N^{\star}$(1710) becomes dominant. Since the excitation of
$N^{\star}$-resonances can only follow an exchange of a non-strange
meson, it can be concluded that there is no dominant kaon
exchange contributing to the reaction. Note that even at 3.30 GeV/c the
$p\Lambda$-FSI has a significant influence on particular regions of
the measured Dalitz plots.

As a next step hyperon production in $pn$-reactions will be studied at
COSY-TOF. First test measurements  showed that events of the
type $pn(p){\to} K^0\Lambda p(p)$ can be identified (first results
from ANKE on the reaction $pn{\to} K^+X$ are presented in
Sect.~\ref{sec:pn}).  Moreover, the use of a polarized beam will allow
one to extract polarization observables. In this context a special
motivation comes from the DISTO experiment where in the reaction $\vec
pp{\to} K^+\vec{\Lambda} p$ the polarization transfer coefficient
$D_{YY}$ has been measured to be strongly negative~\cite{disto},
pointing at a $K$-exchange dominance~\cite{laget_tof}. COSY-TOF allows
one to study this quantity closer to threshold with full phase-space
coverage.

\subsection{\boldmath$\Lambda$N final-state interaction}
\label{sec:YN-fsi}
Although of very high interest, at present little is known about the
strength of the hyperon-nucleon interaction (parameterized by the
scattering length) at small energies not to speak of its spin
dependence. The problem is due to the fact that it is practically
impossible to perform low energy scattering experiments with unstable
particles. Thus, in order to determine the scattering length of this
reaction one has to rely on extrapolations from the data to
threshold. A detailed analysis of the world data set for elastic
$\Lambda p$ scattering gave $a_s{=}-1.8^{+2.3}_{-4.2}\ \mbox{fm and }
a_t{=}-1.6 ^{+1.1}_{-0.8} \ \mbox{fm}$ for the spin-singlet and spin-triplet
scattering lengths, respectively, where the errors are strongly
correlated~\cite{alexander}.

An alternative are production experiments with a hyperon-nucleon pair
in the final state. Then, the scattering parameters are to be
determined from the impact of their FSI on the invariant mass spectra.
In Ref.~\cite{achot} a method has been proposed that allows one to
extract the scattering lengths from the production data directly.  In
particular, an integral representation for the $\Lambda N$ scattering
lengths in terms of a differential cross section of reactions with
large momentum transfer such as $pp{\to} K^+p\Lambda$ or $\gamma
d{\to} K^+n\Lambda$ has been derived.  This formula should enable the
determination of the scattering lengths to a theoretical uncertainty
of about 0.3~fm.

Up to date, all experiments for $\Lambda N$ production in $pp$
collisions were performed unpolarized and thus contain contributions
from both spin triplet as well as spin singlet final states with an
unknown ratio. However, polarization measurements will allow one to
disentangle the different spin states, as is discussed in detail in
Ref.~\cite{achot}. The corresponding experiments can be performed at
COSY.

\section{\boldmath A note on the $\Theta^+$ pentaquark state}
\label{sec:theta+}
The recent discovery of the $\Theta^+$ baryon with strangeness $S {=}
+1$, mass $m_{\Theta^+}\sim1.54$ GeV/c$^2$ and width
$\Gamma<25$~MeV/c$^2$ in various experiments (see {\em e.g.\/}
Ref.~\cite{hep-ph-0312236} and references therein) has triggered an
intensive investigation of exotic resonances ({\em i.e.} non 3$q$
states). After the existence of $\Theta^+$ seems to be confirmed
experimentally its quantum numbers, like spin or parity, have to be
determined.

At COSY the ANKE and COSY-TOF facilities can be used for measurements
of $\Theta^+$-production in hadronic interactions. Since both cannot
detect photons the relevant reaction channels are:
\begin{equation}
pp\to \Sigma^+\Theta^+ \to \Sigma^+\left[pK^0\right] \to
  \left({p\pi^0 \atop n\pi^+}\right)\left[p(\pi^+\pi^-)\right]\ .
\end{equation}
This implies $K^0$ identification by the $(\pi^+\pi^-)$ invariant mass
and the $\Sigma^+$ by the $[p(\pi^+\pi^-)]$ missing mass and asking
for an additional proton ($\Sigma^+\to p\pi^0$) or an additional
positive pion ($\Sigma^+\to n\pi^+$).  For the candidate events
$\{\Sigma^+,K^0\}$ the invariant mass of the $[pK^0]$ subsystem has to
be reconstructed.

At COSY-TOF an about 5$\sigma$ signal in the $K^0p$ invariant mass
distribution has, in fact, been observed at a beam momentum of 2.95
GeV/c. The very preliminary cross section estimate is of the order of
a few hundred nb.  The width of the peak is about 25 MeV/c$^2$,
corresponding to the experimental resolution~\cite{eyrich}. At ANKE a
proposal~\cite{anke-theta} has been accepted to measure the reactions
$pp\to K^0 p \Lambda \pi^+$, $pp\to K^0 p \Sigma^+$, and $pn\to K^0 p
\Lambda$ at maximum COSY energy. These measurements will be carried
out in spring 2004.

In recent papers,  Thomas {\em et al.}~\cite{thomas} and Hanhart
{\em et al.}~\cite{hep-ph-0312236} have emphasized that the parity
of the $\Theta^+$ can be determined from polarization observables of
the reaction $\vec p\vec p \to \Sigma^+\Theta^+$ near the production
threshold.  In particular, the sign of the spin correlation
coefficient $A_{xx}$ yields the negative of the parity of the
$\Theta^+$~\cite{hep-ph-0312236}. Such measurements can, in principle,
be carried out at ANKE and COSY-TOF.

\section{\boldmath Investigation of $K^+$-production on neutrons  with ANKE}
\label{sec:pn}
Data on the $K^+$-production cross section from $pn$ interactions in
the close-to-threshold regime are not available yet. This quantity is,
for example, crucial for the theoretical description of $pA$ and $AA$
data since it has to be used as an input parameter for corresponding
model calculations. Predictions for the ratio $\sigma_{n} /
\sigma_{p}$ range from one to six, depending on the underlying model
assumptions: Pirou\'e and Smith~\cite{Piroue} proposed that there is
no difference between $K^+$ production on the neutron and proton,
whereas the analysis by Tsushima {\em et al.}~\cite{Tsushima} yields
$\sigma_{n} / \sigma_{p}\sim 2$ for the total production cross
sections. F\"aldt and Wilkin~\cite{Wilkin} draw an analogy between
$K^+$- and $\eta$-meson production and give a ratio of six for the
ratio between production on the neutron and proton.

\begin{figure}[hbt]
  \centering \vspace*{-8mm}
  \resizebox{7cm}{4.2cm}{\includegraphics[scale=1.]{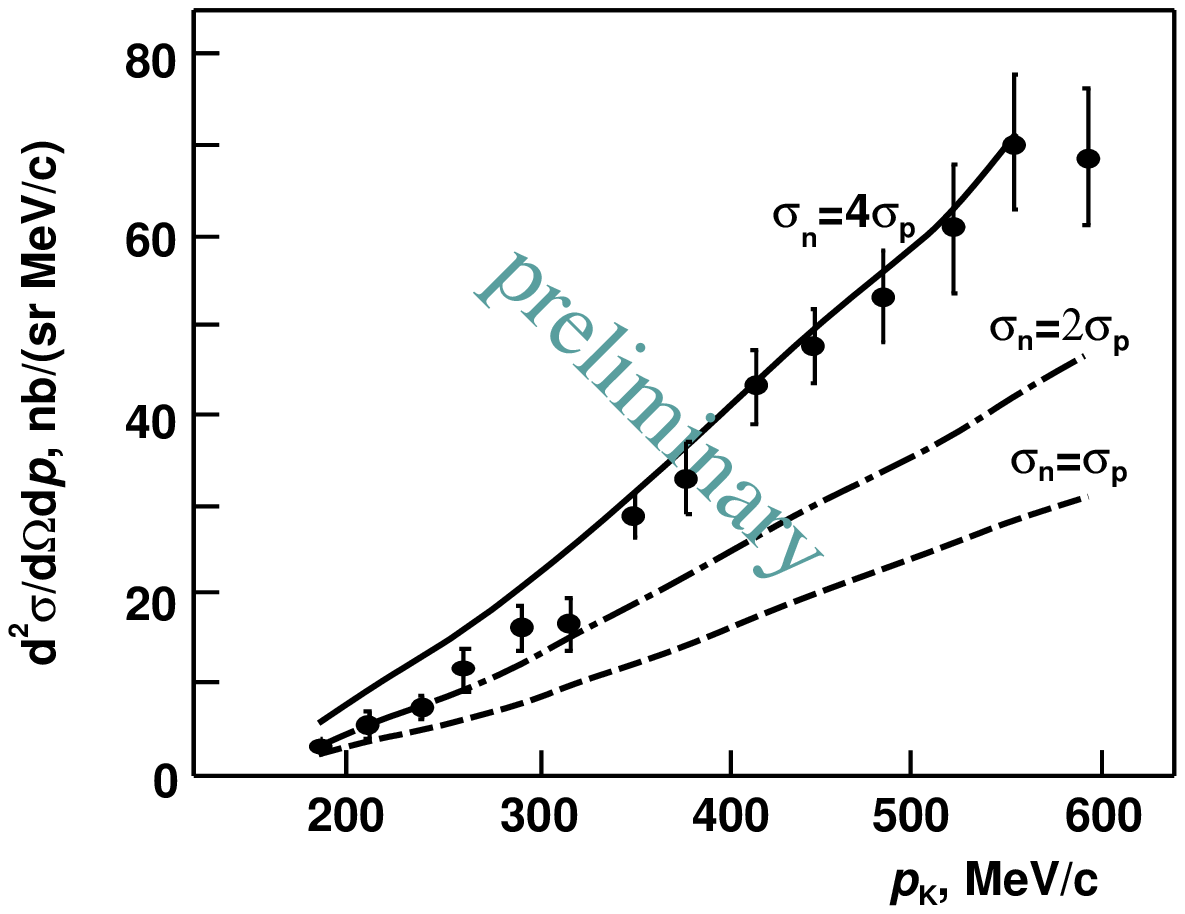}}
  \resizebox{12cm}{3.7cm}{\includegraphics[scale=1.]{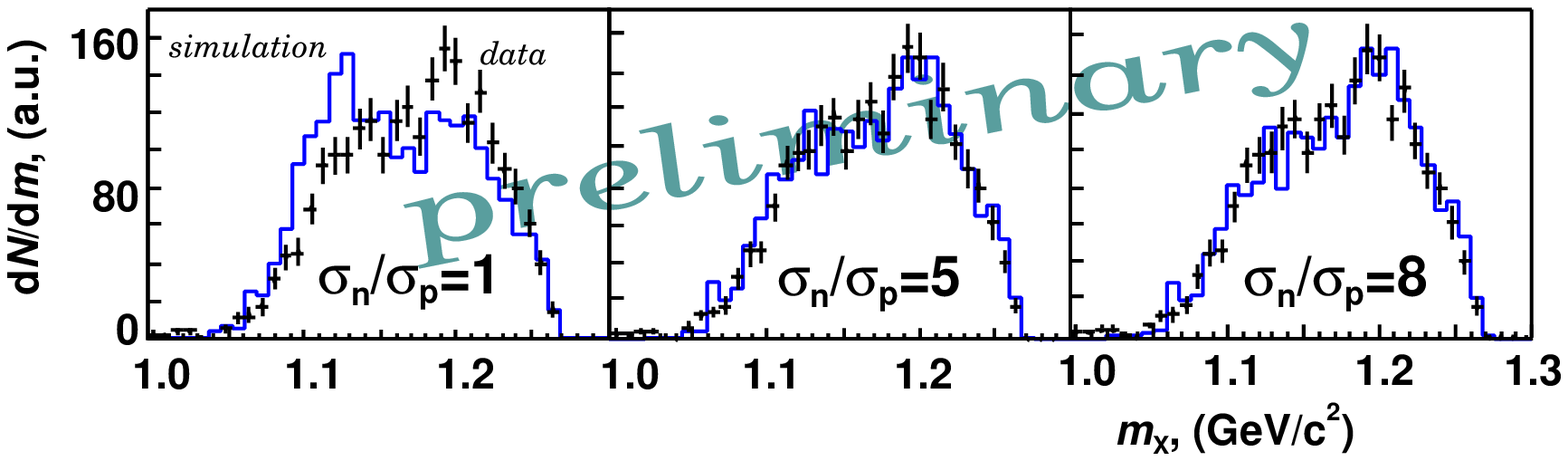}}
  \vspace*{-12mm} 
  \caption{Upper: Double differential $p\mathrm{D}\to K^+X$ cross
  section at 2.83 GeV/c in comparison with model calculations using
  different ratios $\sigma_n/\sigma_p$
  (lines)~\cite{anke_inclusive_K,MazurianLakes}. The overall
  systematic uncertainty from the luminosity normalization of 20\% is
  not included in the error bars. Lower: Missing mass $m_X$ for
  $p\mathrm{D}\to K^+pX(p_{\mathrm s})$ events (with an undetected
  ``spectator'' proton $p_{\mathrm s}$) at $p{=}2.83$ GeV/c in
  comparison with the phase-space calculations.}
  \label{fig:pd2K+X}
\end{figure}

$K^+$-production in $p$D interactions has been measured with ANKE
$p{=}2.60$ and 2.83
GeV/c~\cite{anke_inclusive_K,MazurianLakes}. Figure
\ref{fig:pd2K+X} shows the $K^+$-momentum spectrum for the
higher beam momentum. Based on the assumption that the $K^+$-production
cross section is governed by the sum of the elementary $pp$ and the
$pn$ cross sections, the spectra have been analyzed in a simple
phase-space approach, assuming $\sigma_{\mathrm D} = \sigma_{p} +
\sigma_{n}$.
In Fig.\ \ref{fig:pd2K+X} the resulting momentum spectra are shown
based on the approaches from Ref.~\cite{Piroue} (dashed line
labeled by ``$\sigma_n {=} \sigma_p$'') and Ref.~\cite{Tsushima}
(dash-dotted line, ``$\sigma_n {=} 2\sigma_p$'').

The apparent difference between the calculated and measured cross
sections can be due to the fact that the ratio $\sigma_n/\sigma_p$
differs from those given in Refs.~\cite{Piroue,Tsushima}.  Thus the
simulations were also made keeping the relative weights of the
individual $pp$ and $pn$ channels constant (as given by
Ref.~\cite{Tsushima}) but treating the ratio of the sum of these two
contributions, {\em i.e.\/} $\sigma_{n} / \sigma_{p}$, as a free
parameter.  The best agreement between data and calculations is
obtained for $\sigma_{n} / \sigma_{p} \sim 3$ at 2.60 GeV/c and
$\sigma_{n} / \sigma_{p} \sim 4$ at 2.83 GeV/c (solid line in upper
part of Fig.\ \ref{fig:pd2K+X}).

The resulting large value of $\sigma_{n} / \sigma_{p}$ from the
inclusive spectra is supported by the analysis of missing-mass spectra
from $p\mathrm{D}\to K^+pX(p_{\mathrm s})$ events collected during the
same beam time. The spectrum for $2.83$ GeV/c is also shown in Fig.\
\ref{fig:pd2K+X} and is compared with the result of the Monte-Carlo
simulations, again for different ratios $\sigma_{n} / \sigma_{p}$. The
best agreement between data and simulations is obtained for
$\sigma_{n} / \sigma_{p} \sim (4-5)$.

\section{\boldmath$K\bar K$-pair production at ANKE, COSY-11 and MOMO}
A primary goal of hadronic physics is to understand the structure of
mesons and baryons, their production and decays, in terms of quarks
and gluons. The non-perturbative character of the underlying theory
--- Quantum Chromo Dynamics (QCD) --- hinders straight forward
calculations. QCD can be treated explicitly in the low
momen\-tum-transfer regime using lattice techniques~\cite{lattice},
which are, however, not yet in the position to make quantitative
statements about light scalar states ($J^P{=}0^+$). Alternatively, QCD
inspired models, which use effective degrees of freedom, are to be
used. The constituent quark model is one of the most successful in
this respect (see e.g.\ Ref.~\cite{quarkmodel}). This approach treats
the lightest scalar resonances $a_0/f_0$(980) as conventional
$q\bar{q}$ states. However, they have also been identified with
$K\bar{K}$ molecules~\cite{KK_molecules} or compact
$qq$-$\bar{q}\bar{q}$ states \cite{4q_states}. It has even been
suggested that at masses below 1.0 GeV/c$^2$ a full nonet of 4-quark states
might exist \cite{4q_nonet}. Such possible deviations from the minimal
quark model have a parallel in the baryon sector, where the above
mentioned $\Theta^+$ state requires at least five quarks.

The existing data are insufficient to conclude on the structure of the
light scalars and additional observables are urgently called for. In
this context the charge-symmetry breaking (CSB) $a_0$-$f_0$ mixing
plays an exceptional role since it is sensitive to the overlap of the
two wave functions. It should be stressed that, although predicted to
be large long ago~\cite{achasov}, this mixing has not been identified
unambiguously in corresponding experiments.

\subsection{First experimental results} 
\label{sec:KK_data}
An experimental program has been started at COSY which aims at
exclusive data on $a_0/f_0$ production close to the $K\bar{K}$
threshold from $pp$~\cite{cosy-11-KK,a+_proposal}, $pn$
\cite{a0f0_proposal}, $pd$~\cite{a0f0_proposal,momo-KK} and $dd$ 
\cite{css2002,dd_proposal} interactions --- i.e.\ different isospin 
combinations in the initial state. The reactions $pp {\to} ppK^+K^-$
and $pd {\to} ^3\mathrm{He}\, K^+K^-$ have been measured at COSY-11
\cite{cosy-11-KK,cosy-11-a0f0} and MOMO~\cite{momo-KK}, respectively, at 
excitation energies up to $Q=56$~MeV above the $K\bar K$
threshold. However, mainly due to the lack of precise angular
distributions, the contribution of the $a_0/f_0$ to $K\bar K$
production remains unclear for these reactions.

\begin{figure}[ht]
  \centering \vspace*{-5mm}
  \resizebox{9.5cm}{8.5cm}{\includegraphics[scale=1]{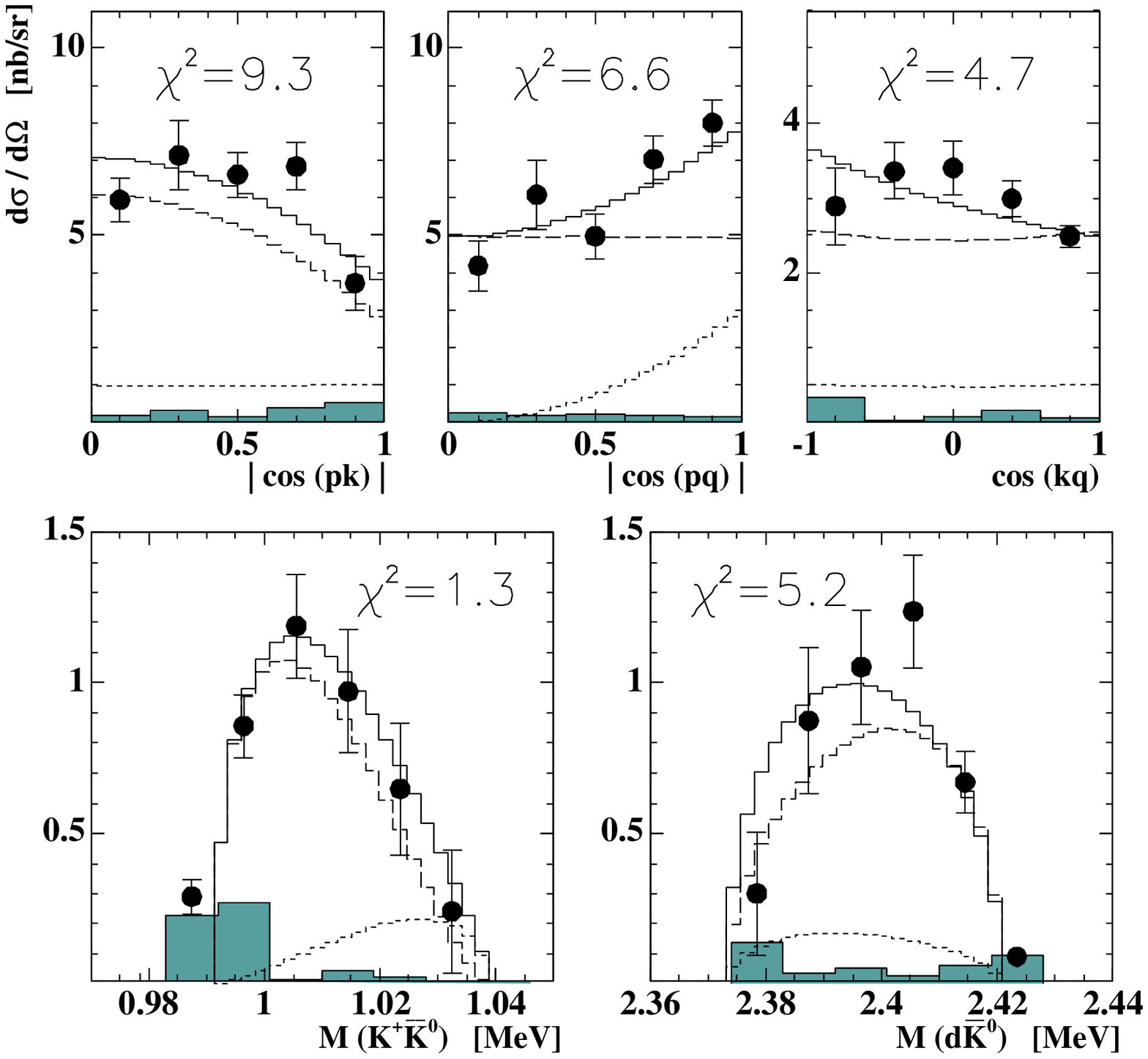}}
  \resizebox{6cm}{!}{\includegraphics[scale=1]{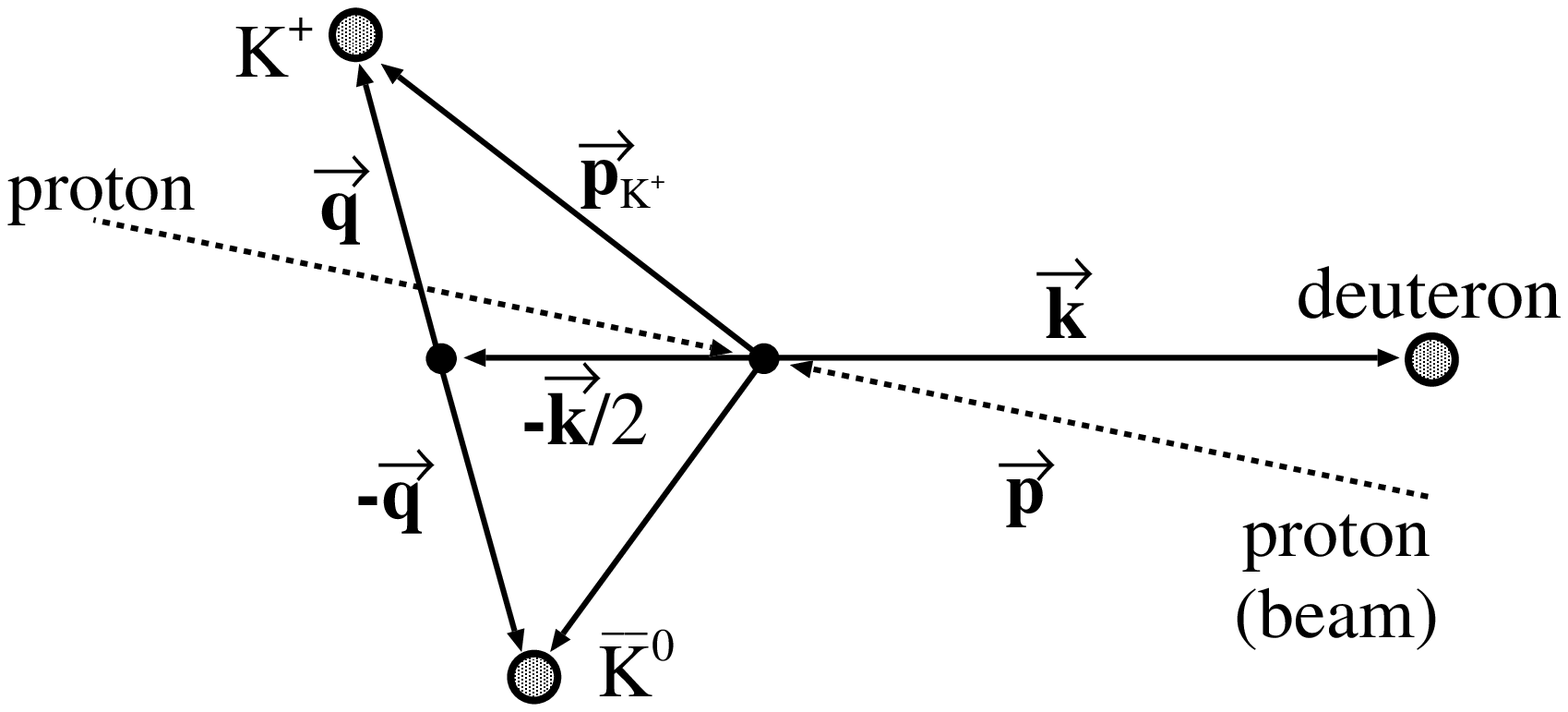}}
  \vspace*{-10mm}

  \caption{ANKE data for the reaction $p(3.46\, \mathrm{GeV/c})p\to
  dK^+\bar{K}^0$~\cite{a+_PRL}. The shaded areas correspond to the
  systematic uncertainties of the acceptance correction. The dashed
  (dotted) line corresponds to $K^+\bar{K}^0$-production in a relative
  $S$-($P$-) wave and the solid line is the sum of both
  contributions. For definition of the vectors $p$, $q$ and $k$ in
  the cms of the reaction $pp {\to} dK^+\bar{K^0}$ see right hand part
  of the figure.  Angular distributions with respect to the beam
  direction $\vec{p}$ have to be symmetric around $90^\circ$ since the
  two protons in the entrance channel are indistinguishable.}
  \label{fig:pp2dKKbar}
\end{figure}

At ANKE, the reaction $pp {\to} dK^+\bar{K^0}$ has been measured
exclusively (by reconstructing the $\bar{K^0}$ from the measured
$dK^+$ missing mass) at beam momenta of $p{=}3.46$ and 3.65 GeV/c
($Q{=}46$ and 103 MeV). These measurements crucially depend on the
high luminosities achievable with internal targets, the large
acceptance of ANKE for close-to-threshold reactions, and the excellent
kaon identification with the ANKE detectors. The obtained differential
spectra for the lower beam momentum are shown in Fig.\
\ref{fig:pp2dKKbar}~\cite{a+_PRL}.

The background of misidentified events in the spectra of Fig.\
\ref{fig:pp2dKKbar} is less than 10\% which is crucial for the 
partial-wave analysis. This analysis reveals that the $K^+\bar{K}^0$
pairs are mainly (83\%) produced in a relative $S$-wave (dashed line
in Fig.\ \ref{fig:pp2dKKbar}), which has been interpreted in terms of
dominant $a_0^+$-resonance production~\cite{a+_PRL}. 

Data on the reaction $p(3.46\, \mathrm{GeV/c}) p\to d \pi^+X$ have
been obtained at ANKE in parallel to the kaon data.  In contrast to
the latter, where the spectra are almost background free, the $pp\to
d\pi^+\eta$ signal is on top of a huge multi-pion background. This
makes the analysis of this channel much more demanding and at present
even model dependent~\cite{a+_pieta}. A total cross section of
$\sigma(pp\to d\pi^+\eta)\sim 4.6\, \mu$b has been extracted from the
data with a resonant contribution of $\sigma(pp\to da_0^+\to
d\pi^+\eta)\sim 1.1\, \mu$b~\cite{a+_pieta}. Together with the cross
section for the $a_0^+\to K^+\bar{K}^0$ channel this yields a
branching ratio of $B.R.(KK/\pi\eta)\sim(0.029{\pm}
0.005_\mathrm{stat} {\pm} 0.02_\mathrm{syst})$ which is in reasonable
agreement with model calculations $B.R.(KK/\pi\eta) \sim
0.04$~\cite{grishina_anke-ws}\footnote{Note that the branching ratio
at beam momenta close to the $K\bar K$ threshold strongly depends on
the $Q$ value of the reaction due to phase-space cuts at the upper
edge of the mass distributions.} for this beam momentum. This
confirms the interpretation of dominant resonant $K^+\bar{K^0}$
production via the $a_0^+$ from above.

The data at $Q{=}103$~MeV are still being analyzed. As the next step a
measurement of the reaction $pn\to dK^+K^-$ at $Q\sim100$ MeV will be
performed at ANKE in Feb.\ 2004. The results of a similar experiment
on the reaction $pn{\to}K^+X$ --- demonstrating the feasibility of
such measurements with a D$_2$ target --- are described in Sect.\
\ref{sec:pn}. According to our cross-section estimates a measurement
of the reaction $dd\to\alpha K^+K^-$ should be feasible within few
weeks of beam time and is foreseen for winter
2004/05~\cite{css2002,dd_proposal,dd_cross-section-est}.

\subsection{Planned investigations of light scalar mesons} 
The experimental results on $a_0^+$ production in $pp$ interactions
can also be regarded as a successful feasibility test for a long-term
experimental program with the final goal to determine the
charge-symmetry breaking $a_0$-$f_0$ mixing amplitude. These
measurements will require the use of a photon detector which is not
yet available at COSY. However, it is planned to relocate the WASA
detector~\cite{wasa} from its current location at CELSIUS/TSL to COSY
in summer 2005, which will then make these experiments feasible.

Both, the $a_0^0$- and the $f_0$-resonances can decay into $K^+K^-$
and $K_SK_S$, whereas in the non-strange sector the decays are into
different final states according to their isospin, $a_0^\pm{\to}
\pi^\pm\eta$, $a_0^0{\to} \pi^0\eta$ and $f_0{\to} \pi^0\pi^0$ or 
$\pi^+\pi^-$.  Thus, only the non-strange decay channels have defined
isospin and allow one to discriminate between the two mesons. It is
also only by measuring the non-strange decay channels that CSB can be
investigated.  Such measurements can be carried out with WASA at COSY
for active $\pi^0$- or $\eta$-meson identification, while the strange
decay channels $a_0/f_0{\to} K_SK_S$ should be measured in parallel.

Since it is possible to manipulate the initial isospin of purely
hadronic reactions one can identify observables that vanish in the
absence of CSB~\cite{miller,ANKE_WS}. The idea behind the proposed
experiments is the same as behind recent measurements of CSB effects
in the reactions $np{\to} d\pi^0$~\cite{opper} and $dd{\to}\alpha\pi^0$
\cite{stephenson}. However, the interpretation of the signal from the 
scalar mesons is much simpler as compared to the pion case. Since the
$a_0$ and the $f_0$ are rather narrow overlapping resonances, the
$a_0$-$f_0$ mixing in the final state is enhanced by more than an
order of magnitude compared to CSB in the production operator (i.e.\
``direct'' CSB violating $dd{\to} \alpha a_0$ production) and should,
e.g., give the dominant contribution to the CSB effect via the
reaction chain $dd{\to} \alpha f_0(I{=}0) {\to} \alpha a_0^0(I{=}1)
{\to} \alpha (\pi^0\eta)$~\cite{CH}. This reaction seems to be most
promising for the extraction of CSB effects, since the initial
deuterons and the $\alpha$ particle in the final state have isospin
$I{=}0$ (``isospin filter''). Thus, any observation of $\pi^0\eta$
production in this particular channel is a direct indication of CSB
and can give information about the $a_0$-$f_0$ mixing
amplitude~\cite{CH}.

In analogy with the measurement of CSB effects in the reaction $np{\to}
d\pi^0$, it has been predicted that the measurement of angular
asymmetries (i.e.\ forward-backward asymmetry in the $da_0$ c.m.s.) 
can give information about the $a_0$-$f_0$ mixing
\cite{a0_f0-mixing_PLB,tarasov,kudr}. It was stressed in Ref.\ 
\cite{tarasov} that --- in contrast to the $np{\to} d\pi^0$ experiment 
where the forward-backward asymmetry was found to be as small as 0.17\%
\cite{opper} --- the reaction $pn{\to} d\pi^0\eta$ is subject to a
kinematical enhancement.  As a consequence, the effect is predicted to
be significantly larger in the $a_0$/$f_0$ case. The numbers range
from some 10\%~\cite{tarasov} to factors of a few
\cite{a0_f0-mixing_PLB} and, thus, should easily be observable. It has 
been pointed out in Ref.~\cite{kudr} that the analyzing power of the
reaction $\vec p n{\to} d \pi^0 \eta$ also carries information about the
$a_0$-$f_0$ mixing amplitude. This quantity can be measured at COSY as
well.

\section{Acknowledgments}
The author is grateful to C.~Hanhart for contributing to Sect.\
\ref{sec:YN-fsi}, as well as to W.~Eyrich, W.~Oelert and H.~Str\"oher 
for carefully reading the manuscript.

\end{document}